\newcommand{\SigmaAv}[0]{{\overline{\sigma}}}
\newcommand{\SigmaDom}[0]{\sigma_{\rm{m}}}
\newcommand{\SigmaProt}[0]{\sigma_{\rm{p}}}
\newcommand{\Dfst}[2]{\frac{\partial #1}{\partial #2}}
\newcommand{\Dscd}[2]{\frac{\partial ^2 #1}{\partial #2 ^2}}
\newcommand{\defEquation}[1] {Eq.\ \ref{#1}}
\newcommand{\defEquations}[2] {Eqs.\ \ref{#1} and \ref{#2}}
\newcommand{\defFigure}[1] {Fig.\ \ref{#1}}
\newcommand{\defFigures}[2] {Figs.\ \ref{#1} and \ref{#2}}
\newcommand{\defBindingDomain}[0] {adsorption domain}
\newcommand{\defTotalFBare}[0] {F^{(0)}} 
\newcommand{\defMuBare}[0]{\mu_i^{(0)}} 
\newcommand{\defTotalFDom}[0]{F}
\newcommand{\defEsFDom}[0] {F_{\rm{es}}}
\newcommand{\defEsFBare}[0]{F_{\rm{es}}^{(0)}}
\newcommand{\defMuDom}[0]{\mu_i} 
\newcommand{\defPIP}[0]{PIP$_2$}
\newcommand{\defFAMARCKS}[0]{FA-MARCKS(151-175)}
\newcommand{\defLys}[0]{(Lys)$_{13}$}
\newcommand{\defRefManuscript}[0]{manuscript submitted for
  publication}

\newcommand{\we}[0]{we}
\newcommand{\We}[0]{We}
\newcommand{\Our}[0]{Our}
\newcommand{\our}[0]{our}
\newcommand{\ourthis}[0]{our}
\newcommand{\us}[0]{us}
\newcommand{\envIncludeEps}[2]{\includegraphics[width=#2]{#1}}
\newcommand{\envPageBreak}[0]{\newpage} 

\newcommand{\envSubSection}[1]{\subsection{#1}}
\newcommand{\envSection}[1]{\subsection{#1}}
\newcommand{\envAbstractBegin}[0]{\envSection{ABSTRACT}}
\newcommand{\envAbstractEnd}[0]{\envPageBreak{}}
\newcommand{\envPictureLocation}[0]{tbp}
\newcommand{\envCaptionFont}[0]{\footnotesize}

\newcommand{\envMakeTitle}[0]
{
  \author{Emir Haleva,$^*$ Nir Ben-Tal,$^*$ and Haim Diamant$^\#$\\ 
    \footnotesize $^*$ Department of Biochemistry, George S. Wise
    Faculty of Life Sciences, Tel Aviv University, Ramat\\ 
    \footnotesize Aviv 69978, Israel. $^\#$ School Of Chemistry,
    Raymond and Beverly Sackler Faculty of Exact Sciences,\\ 
    \footnotesize Tel Aviv University, Ramat Aviv 69978, Israel.}
  \maketitle{} 
  \noindent \textit{Correspondence}: Haim Diamant, School Of
  Chemistry, Raymond and Beverly Sackler Faculty of Exact Sciences,
  Tel Aviv University, Ramat Aviv 69978, Israel; Tel.:
  +972-3-640-6967; Fax: +972-3-640-9293; E-mail:
  hdiamant@tau.ac.il \\ \\ 
  \textit{Running Title}: Protein-Induced Phospholipid
  Redistribution\\ 
  \\ \textit{Keywords}: theory, double-layer interaction,
  protein--membrane interactions, MARCKS, \defPIP{} 
  \setcounter{secnumdepth}{0} 
  }

\newcommand{\envMakeAcknowledgments}[0]{
  \vspace{0.3in} \noindent \footnotesize 
  We thank S.\ McLaughlin and
  D.\ Murray for providing us with draft copies of their papers prior
  to submission, and we are grateful to S.\ McLaughlin for comments on
  the manuscript and for valuable discussions.  We benefited from
  discussions with D.\ Andelman, Y.\ Burak, M.\ Gutman, and M.\
  Kozlov.\ H.\ D.\ acknowledges support from the Israeli Council of
  Higher Education (Alon Fellowship).  \normalsize }
  \newcommand{\envThesisRemark}[1]{#1}
\newcommand{\envArticleRemark}[1]{#1}

\documentclass[12pt]{article}

\newcommand{\envDraft}[0]{ 
  \renewcommand{\baselinestretch}{1}
  \renewcommand{\envThesisRemark}[1]{}
  \renewcommand{\envPictureLocation}[0]{tbh}
  \usepackage{my_refernce_ind}
  \renewcommand{\envAbstractEnd}{}
  }

\renewcommand{\baselinestretch}{2}

\envDraft{}

\usepackage{graphicx}
\usepackage{amssymb}
\usepackage{subfigure} 

\newcommand{\defRefGambhir}[0]{Gambhir et al., \defRefManuscript{}}
\newcommand{\defRefDiana}[0]{Wang et al., \defRefManuscript{}} 

\begin{document}

\title{\envThesisRemark{\vspace{-0.9in}TEL-AVIV UNIVERSITY\\GEORGE S. WISE FACULTY OF
    LIFE SCIENCES\\GRADUATE SCHOOL\\ \vspace{0.3in}}Increased
  Concentration of Polyvalent Phospholipids in the Adsorption Domain
  of a Charged Protein} 
\envMakeTitle{}


\envAbstractBegin{} \We{} studied the adsorption of a charged protein
onto an oppositely charged membrane, composed of mobile phospholipids
of differing valence, using a statistical-thermodynamical approach.  A
two-block model was employed, one block corresponding to the
protein-affected region on the membrane, referred to as the
\defBindingDomain{}, and the other to the unaffected remainder of the
membrane. \We{} calculated the protein-induced lipid rearrangement in
the \defBindingDomain{} as arising from the interplay between the
electrostatic interactions in the system and the mixing entropy of the
lipids. Equating the electrochemical potentials of the lipids in the
two blocks yields an expression for the relations among the various
lipid fractions in the \defBindingDomain{}, indicating a sensitive
dependence {of lipid fraction on valence.} 
This expression is a
result of the two-block picture but does not depend on further details
of the protein--membrane interaction. \We{} subsequently calculated
the lipid fractions themselves using the Poisson-Boltzmann theory.
\We{} examined the dependence of lipid enrichment, i.e., the ratio
between the lipid fractions inside and outside the
\defBindingDomain{}, on various parameters such as ionic strength and
lipid valence. Maximum enrichment was found for lipid valence of about
$(-3)$ to $(-4)$ in physiological conditions.  \Our{} results are in
qualitative agreement with recent experimental studies on the
interactions between peptides having a domain of basic residues and
membranes containing a small fraction of the polyvalent
phosphatidylinositol 4,5-bisphosphate (PIP$_2$). This study provides
theoretical support for the suggestion that proteins adsorbed onto
membranes through a cluster of basic residues may sequester \defPIP{}
and other polyvalent lipids.

\envThesisRemark{I further examined one of the assumptions of the
  model, which treated the protein and membrane as objects of infinite
  width, by studying the interaction of two impenetrable parallel
  plates of finite width and varying characteristics. It is shown
  that, in this kind of electrostatic interaction, objects such as
  peptides could be treated to a good approximation as having infinite
  width.}

\envAbstractEnd{}

\envSection{INTRODUCTION} 
Some membrane-associated proteins are known to bind to membranes
non-specifically through electrostatic interactions (Murray et al.,
1997; Resh et al., 1999; McLaughlin et al., 2002; Murray et al.,
2002). These interactions result from the attraction between a cluster
of charged residues in the protein and the oppositely charged membrane
lipids. 

{As the charged protein approaches the membrane, it changes the
local membrane composition in its vicinity.} \We{} refer to this
protein-affected region on the membrane as the
\defBindingDomain{}. The lateral fluidity of the membrane allows
oppositely charged lipids to migrate toward the
\defBindingDomain{} in order to minimize the interaction free energy.
Evidence for such redistribution was reported in experimental
(Heimburg et al., 1999; Rauch et al., 2002; Wang et al., 2002) and
theoretical (Harries et al., 1998; 
May et al., 2000; Fleck et al.,
2002; May et al., 2002) studies. 

{Lipid redistribution was observed
also in monolayers (Lee et al., 1994; Lee and McConnell, 1995) and
bilayers (Groves et al., 1997; Groves et al., 1998) that were
subjected to external electric fields.}

Local changes in lipid concentration may have biological significance.
For example, \defPIP{}, a polyvalent phospholipid with valence in the
range $-3$ to $-5$ (Toner et al., 1988; McLaughlin et al., 2002; Rauch
et al., 2002; Wang et al., 2002) participates in signal transduction
(Czech, 2000; Payrastre et al., 2001; Simonsen et al., 2001). Its
average fraction in plasma membranes is very low, about 1\%, and it is
known to be concentrated in specific regions of the membrane (Liu et
al., 1998; Stauffer et al., 1998). The \defPIP{} lipid serves as a
substrate for Phospholipase C (PLC), which cleaves it to two secondary
messengers (Katan and Williams, 1997). Another component, the
Myristoylated Alanine-rich C Kinase Substrate protein (MARCKS),
containing an amino acid segment of 13 basic and no acidic residues
(Blackshear, 1993; McLaughlin and Aderem 1995), is believed to form a
\defPIP{} `reservoir' in its \defBindingDomain{}. As long as PIP$_2$
is concentrated in the MARCKS \defBindingDomain{}, PLC is inhibited
and cannot catalyze the PIP$_2$ hydrolysis (Wang et al., 2001). It is
assumed that upon demand, by phosphorylating the MARCKS effector
segment, these lipids are freed for signaling (McLaughlin et al.,
2002). Thus the capability of MARCKS to sequester PIP$_2$ potentially
affects intracellular signaling.

Lipids of various valences are attracted to the \defBindingDomain{} to
different extents. This electrostatically-induced enrichment is
partially balanced by entropy effects that favor homogeneous lipid
distribution. Theoretical studies of bilayers composed of neutral and
monovalent lipids, where the lipid mobility was taken into account,
showed that the formation of a charged lipid domain due to the
adsorbed protein is energetically favorable and outweighs entropy
effects (e.g., Heimburg et al., 1999; 
May et al., 2000).  Recently, Fleck et al. (2002) presented a detailed
formulation for the interactions between charged objects and a
membrane composed of lipids of various valences. These studies
demonstrate the importance of lipid redistribution in the
thermodynamics of protein--membrane adsorption.

Based on the findings of these theoretical studies,
\envArticleRemark{we focus in this work} \envThesisRemark{this work is
  focused }on a simplified model for the redistribution of
different-valence lipids in the \defBindingDomain{} of a charged
protein. \We{} start by deriving an expression that relates the
fractions of the various lipids in the \defBindingDomain{} to their
values in the unperturbed membrane. The general expression is
restricted, however, to \emph{relations} between concentrations of
different lipids. To obtain the actual concentration values for each
lipid type, \we{} use the Poisson-Boltzmann (PB) theory (e.g.,
Andelman, 1995).  \envThesisRemark{The full nonlinear
  Poisson-Boltzmann (NLPB) case is analyzed, and subsequently a
  simplified solution is presented assuming weak electrostatics (the
  Linear Poisson-Boltzmann approximation, LPB).}
The model is then applied to peptide segments, such as a polylysine
chain, interacting with a membrane composed of uncharged, monovalent,
and trivalent lipids, corresponding to zwiterionic PC, PS, and
\defPIP{}, respectively. Finally, the model is evaluated and the
biological implications of its results are discussed.

\envSection{MODEL}

\begin{figure}[\envPictureLocation{}]
  \center{
    \envIncludeEps{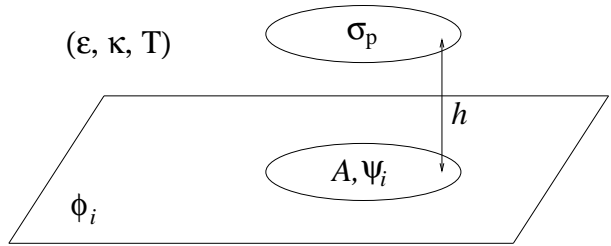}{3.15in}
    \caption[Schematic view of the system]{\envCaptionFont{}Schematic
      view of the protein--membrane system. The protein region
      interacting with the membrane is modeled as a planar surface of
      charge density $\SigmaProt$, hovering parallel to the membrane
      at a distance $h$.  The unperturbed membrane is composed of
      various lipids, each having valence $z_i$ and mole fraction
      $\phi_i$. The interaction region on the membrane, the
      \defBindingDomain{}, has area $A$ and lipid fractions $\psi_i$.
      The whole system is embedded in an ionic solution characterized
      by a dielectric constant $\epsilon$, Debye screening length
      $\kappa^{-1}$ and temperature $T$. }
    \label{figMembraneModel}
}
\end{figure}
  
A schematic view of the protein--membrane system under study is
provided in \defFigure{figMembraneModel}. In the model, the membrane
is considered to be an infinite surface composed of $k$ phospholipid
species.  Each phospholipid type $i$ is ascribed a fraction (in the
unperturbed membrane) $\phi_i$ and a fixed valence $z_i$. (\We{} do
not consider pH-dependent dissociation of charged lipids, which was
found to have a minor effect compared to lipid mobility; Fleck et al.,
2002.)  The indices $i = 0$ and $i = 1$ are assigned to neutral lipids
($z_0 =0$) and monovalent anionic ones ($z_1 = -1$), respectively,
which are always present in biological membranes. For simplicity, all
phospholipids are ascribed the same headgroup area $a$. (In a more
detailed model, one can treat different headgroup areas; Andelman et
al., 1994.)

\We{} assume that the \defBindingDomain{} is a ``patch'' of finite
area $A$ and uniform charge density $\SigmaDom$, whose size is much
larger than the Debye screening length of the solution $\kappa^{-1}$,
i.e., $\kappa^2 A \gg 1$. The screening provided by the surrounding
ionic solution ensures a cutoff for the effect of the adsorbed protein
on the membrane, justifying the finite-area assumption. The uniform
charge density, employed merely for simplicity, can be thought of as
an effective or average domain charge density. \We{} thus neglect
effects related to charge discreteness. (This assumption becomes
invalid in certain circumstances; \we{} shall return to it in the
Discussion.)  The membrane outside the domain serves as a large
reservoir, assumed to be unaffected by the protein.  Thus, in this
study {(similar to the description of May et al., 2002,
who treated many proteins)} \we{} regard the membrane as
composed of two blocks, the protein-affected \defBindingDomain{}, with
lipid fractions $\psi_i$, and the unaffected remainder of the
membrane, with fractions $\phi_i$.  

{Both blocks are assumed to contain
enough molecules to be considered, to a good approximation, as
macroscopic phases. (A treatment of finite-size effects can be found
in May et al., 2002.)} 
The charge densities $\SigmaAv$ and $\SigmaDom$,
in the protein-free and protein-bound regions, respectively, are
defined as:
\begin{equation}
  \label{eqRelationSigma}
  \SigmaAv = \frac{e}{a} \sum_{i}z_i \phi_i  \; \; , 
  \; \; \SigmaDom = \frac{e}{a} \sum_{i}z_i \psi_i
\end{equation}
where $e$ is the elementary charge.

The free energy per unit area of the bare (protein-free) membrane,
$\defTotalFBare{}$, is
\begin{equation}
  \label{eqGBare}
  \defTotalFBare{} = \frac{T}{a} \sum_{i}{\phi_i \ln \phi_i} +  \defEsFBare(\SigmaAv)
\end{equation}
The first term is the mixing entropy contribution, where $T$ is the
temperature in energy units (taking the Boltzmann constant as unity).

{We neglect short-range, non-electrostatic
interactions between lipid molecules (May et al., 2002) except for 
excluded-volume effects.}
 The second term accounts for the electrostatic contribution.
Similarly, the free energy per unit area of the protein-bound domain,
$\defTotalFDom{}$, is
\begin{equation}
  \label{eqGBin}
  \defTotalFDom{} = \frac{T}{a} \sum_{i}{\psi_i \ln \psi_i} + 
  \defEsFDom(\SigmaDom) 
\end{equation}
where $\defEsFDom{}$ accounts for the electrostatic interactions among
the phospholipids and between them and the protein. 
Note that $\defEsFBare{}$
and $\defEsFDom{}$ are functions of $\phi_i$ and $\psi_i$,
respectively, only via the charge densities $\SigmaAv$ and $\SigmaDom$
as defined in
\defEquation{eqRelationSigma}. These free energies can be calculated
using various theories, e.g., the commonly used PB theory (see
Appendix II).  However, at this stage of \our{} formulation, \we{}
need not specify the expressions for $\defEsFBare{}$ and
$\defEsFDom{}$ at all.

The \defBindingDomain{} and the rest of the membrane are at
thermodynamic equilibrium, thus the electrochemical potentials
$\defMuDom{}$ of each lipid type are equal in the two regions. In
addition, the membrane incompressibility adds two constraints:
\begin{equation}
  \label{eqSumEq1}
  \sum_i\phi_i = 1  \; \; , \; \; \sum_{i}\psi_i = 1
\end{equation}
The electrochemical potential of phospholipid $i$ in the protein-free
membrane is
\begin{equation}
  \label{eqMuIBare}
  \defMuBare{} =  a \Dfst{\defTotalFBare}{\phi_i} = 
  T\ln{\frac{\phi_i}{\phi_0}} + z_i e \Dfst{\defEsFBare}{\SigmaAv}
\end{equation}
and similarly, in the \defBindingDomain{},
\begin{equation}
  \label{eqMuI}
  \defMuDom{} =  a \Dfst{\defTotalFDom}{\psi_i} = 
  T \ln{\frac{\psi_i}{\psi_0}} + z_i e \Dfst{\defEsFDom}{\SigmaDom}
\end{equation}
In \defEquations{eqMuIBare}{eqMuI} the dependencies on $\phi_0$ and
$\psi_0$ arise from the incompressibility constraint,
\defEquation{eqSumEq1}. They can be viewed as partial surface
pressures exerted by the uncharged species ($i=0$) due to
excluded-volume effects. (Their surface pressure is equal to
$-T\ln\phi_0$ and $-T\ln\psi_0$, respectively, in the bare membrane
and in the \defBindingDomain{}.) \We{} still have not specified
explicit expressions for $\defEsFBare{}$ and $\defEsFDom$. As a
particular example, one may assume a mean electric potential (as in
PB), having a value $\Psi^{(0)}(0)$ at the bare membrane, and then
$\partial \defEsFBare/ \partial \SigmaDom = \Psi^{(0)}(0)$. If, in
addition, \we{} set $\phi_0 \simeq 1$ (low membrane charge), then the
familiar expression for the electrochemical potential is recovered,
$\defMuBare = T\ln{\phi_i} + z_i e
\Psi^{(0)}(0)$ (and similarly for $\defMuDom$).  Note, however, that
the validity of \envArticleRemark{our} \envThesisRemark{the presented}
formulation is more general than this specific example.

Equating $\defMuBare = \defMuDom$ we get
\begin{equation}
  \label{eqMuCompare}
  \frac{e}{T} \left(\Dfst{\defEsFBare}{\SigmaAv} - 
    \Dfst{\defEsFDom}{\SigmaDom}\right) = 
  \frac{1}{z_i} \ln\frac{\psi_i \phi_0}{\phi_i \psi_0}
\end{equation}
Importantly, the left-hand side of \defEquation{eqMuCompare} is
independent of $i$. \We{} can therefore compare the right-hand side of
the equation for a certain species $i$ with the same expression for
the monovalent species ($i$ = 1). This gives a set of equations
relating the enrichment ratios for the various lipids,
\begin{equation}
  \label{eqRelationEquation}
  \frac{\psi_i}{\phi_i} =
  \left(\frac{\psi_1}{\phi_1}\right)^{z_i/z_1}
  \left(\frac{\phi_0}{\psi_0}\right)^{z_i/z_1 -1}
\end{equation}

\We{} refer to \defEquation{eqRelationEquation} as the \emph{relative
  enrichment equation}. It is a set of $(k - 2)$ equations for every
$i$ $\neq$ 0,1. \envThesisRemark{(Note that $z_1 = -1$ in \ourthis{}
  study.) }The relative enrichment equation is one of the key results
of this work. In the limit of low charge density, $\phi_0, \psi_0
\simeq 1$, and assuming a mean electric potential having the values
$\Psi^{(0)}(0)$ and $\Psi(0)$ at the unperturbed membrane and the
\defBindingDomain{}, respectively, \defEquation{eqRelationEquation}
is directly related to a Boltzmann
relation,
\begin{equation}
  \psi_i \simeq
  \phi_i \exp[-e z_i(\Psi(0)-\Psi^{(0)}(0))/ T]
\label{Boltzmann}
\end{equation}
As derived above, however, the applicability of
\defEquation{eqRelationEquation} is more general; it is more
accurate and is valid for highly charged membranes and beyond the
mean-field approximation. 

{In particular, \defEquation{Boltzmann}
implies that the protein does not perturb the local concentration of
the neutral lipid, $\psi_0/\phi_0\simeq 1$. Yet, as we shall see
below, in the biologically relevant case where the membrane contains a
large fraction of monovalent lipid, the neutral lipid is significantly
depleted from the adsorption domain.  Hence,
\defEquation{eqRelationEquation}, rather than \defEquation{Boltzmann},
will be used throughout our analysis.}

In the absence of protein, the phospholipid composition would not
change and both sides of
\defEquation{eqRelationEquation} are trivially equal to 1.  Another
consequence of \defEquation{eqRelationEquation} is that perturbation
of one lipid fraction necessarily entails perturbation in others. In
cases where the charge density in the \defBindingDomain{} increases
(in absolute value), $|\SigmaDom| > |\SigmaAv|$, neutral phospholipids
will be depleted from the domain to allow the entrance of charged
ones, $\phi_0 / \psi_0 > 1$. The relative enrichment of lipid $i$,
$\psi_i / \phi_i$, is then at least that of the monovalent one raised
to the power $|z_i|$.

As an example, let us consider the binding of a basic protein to a
membrane containing uncharged zwiterionic lipids (e.g., PC),
monovalent lipids (e.g., PS) and trivalent ones (e.g., PIP$_2$).
According to \defEquation{eqRelationEquation}, the polyvalent
enrichment ratio will be stronger than that of the monovalent fraction
by at least a power of $z_i/z_1 = 3$. This is a strong effect. If the
monovalent concentration increases twofold, the trivalent
concentration will increase by a factor of $2^3=8$.  Similarly, a
slight decrease in the negative charge of the \defBindingDomain{}
leads to a significant decrease in polyvalent fraction. 

{This entropy-driven enhancement of high-valence ion concentration is an
example of a more general phenomenon manifest, e.g., in the favorable
accumulation of dissolved polyvalent ions near charged surfaces or
polyelectrolytes (e.g., Rouzina and Bloomfield, 1996).}
Note again
that the relative enrichment equation is independent of the specific
expressions for $\defEsFBare(\overline{\sigma})$ and
$\defEsFDom(\SigmaDom)$. It does not depend explicitly on details such
as the distance between the protein and membrane, or the protein
charge.

Equation \ref{eqRelationEquation} provides us only with a
\emph{relation} between the different fractions $\psi_i$. To calculate
the actual values of $\psi_i$ \we{} need to derive explicit
expressions for the electrochemical potentials $\defMuBare$ and
$\defMuDom$. To this end, \we{} must introduce details of the protein.
It is treated, for simplicity, as a flat surface of uniform charge
density $\SigmaProt$ located at a distance $h$ parallel to the
membrane (see \defFigure{figMembraneModel}). This schematic
description may be relevant to proteins which have a flat cluster of
basic residues facing the membrane at close proximity, while the rest
of the charged residues are further away, screened by the ionic
solution.  \We{} regard the protein--membrane distance $h$ as an
external parameter determined by other interactions (e.g., desolvation
effects), which are not taken into account in \our{} theory. \We{}
further assume that $h$ is small enough ($h \ll \sqrt A$), such that
the induced \defBindingDomain{} on the membrane and the interacting
cluster on the protein can be taken to have roughly the same area $A$.


\We{} apply the commonly used mean-field Poisson-Boltzmann (PB) theory 
(e.g., Andelman, 1995) to calculate the electrochemical potentials
$\defMuBare$ and $\defMuDom$. 

{The applicability of this theory to
solutions containing polyvalent ions is questionable (e.g., Netz,
2001). However, in the systems discussed here the polyvalent ions
(phospholipids) are restricted to the membrane and their mole fraction
is much smaller than that of the monovalent lipids, in accord with
biological conditions. The more mobile salt ions in the aqueous
solution are assumed to be monovalent. Polyvalent ions thus enter the
PB calculation merely as a (minor) contribution to the membrane
surface charge. As such they should not induce strong correlations,
and the PB theory should be applicable. (An exception, where the
polyvalent lipid is the majority charge in the membrane and simple
electrostatic models indeed seem to fail, will be presented in the
Discussion.)}

\We{} derive $\defMuBare$ and $\defMuDom$ using three alternative 
levels of approximation, all of which are discussed in detail in
Appendix I. 

{In presenting the three methods we wish to demonstrate
that even a much simplified approach, involving minimum computation,
still gives useful results.}  
The nonlinear problem (NLPB) resulting
from the PB theory can only be treated numerically.  Subsequently,
\we{} examine a further approximation where the PB expressions are
linearized (LPB).  This approximation is valid when the electrostatic
potential $\Psi$ is much smaller than $T/e$ everywhere (Andelman,
1995).  Although this condition is
not fulfilled in the relevant biological systems, the two
derivations give similar results 

{for reasons that are discussed
in Appendix I.}
The LPB approximation allows us to derive analytical
expressions for $\defMuDom$, yet solutions for the various lipid
fractions (i.e., for the equations $\defMuBare =
\defMuDom$) still cannot be obtained in closed form.

The fact that, in most relevant systems, the electrostatic
interactions dominate over entropy effects (May et al., 2000) led
\us{} to examine one last approximation, in which the
derivation is divided into two stages. In the first, \we{} neglect
entropy when equating $\defMuBare = \defMuDom$. This implies that the
electrostatic potential is uniform along the membrane (i.e., the
membrane behaves as a perfect conductor). The resulting membrane
charge density in the
\defBindingDomain{} is
\begin{equation}
  \label{eqAnalyticSigma}
  \SigmaDom = \frac{\sinh(\kappa h) \SigmaAv - \SigmaProt}{\cosh(\kappa h)}
\end{equation}
In the second stage, this value of $\SigmaDom$ is substituted in
\defEquation{eqRelationSigma}. Equations \ref{eqRelationSigma},
\ref{eqSumEq1}, and \ref{eqRelationEquation} 

{(the latter incorporating
the effect of entropy)} 
thus provide a closed set of $k$ polynomial
equations which can be easily solved for the $k$ lipid fractions
$\psi_i$. \We{} refer to this scheme as the simplified linear
Poisson-Boltzmann method (SLPB).

\envSection{RESULTS} 
To study the effects of protein adsorption on a
mixed membrane, \we{} calculated the lipid fractions for several
representative conditions. \Our{} aim is to examine the redistribution
of different-valence lipids in the \defBindingDomain{} of a
membrane-adsorbed protein as a function of several parameters:
protein--membrane distance, protein charge, and the valence of the
most charged lipid species.  Unless otherwise stated, \we{} use
physiological values for the Debye length ($\kappa^{-1}$ = 10 \AA{}),
temperature (300 K), dielectric constant of water ($\epsilon = 80$),
and lipid headgroup area ($a$ = 70 \AA{}$^2$).  Throughout the text,
the notation 69\%/30\%/1\% uncharged/monovalent/polyvalent is used to
describe the unperturbed membrane composition ($\phi_i$ values). For
consistency, the results presented in this section were all obtained
using the more elaborate NLPB method. 

\begin{figure}[\envPictureLocation{}]
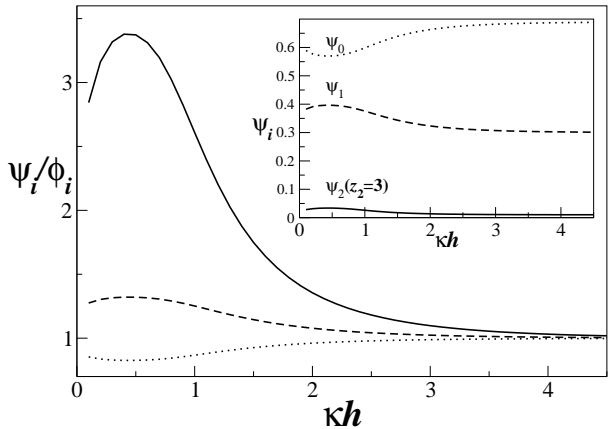

  \center{
    \envIncludeEps{fig2.eps}{3.15in}
    \caption[Protein--membrane distance
    effect]{\envCaptionFont{}Protein--membrane distance effect.
      Enrichment ratios ($\psi_i/\phi_i$) of uncharged (dotted),
      monovalent (dashed) and trivalent (solid) lipid fractions in the
      \defBindingDomain{} of a protein as a function of the
      protein--membrane distance $h$, scaled by the Debye length
      $\kappa^{-1}$. The unperturbed membrane contains 69\%/30\%/1\%
      uncharged/monovalent/trivalent lipid fractions. The charge
      density of the protein is 1.3-fold that of the unperturbed
      membrane and of opposite sign.  The inset shows the actual lipid
      fractions. Although quite similar in charge density to the
      membrane, the protein causes a marked change in lipid
      composition.  Note the relatively large increase in the
      trivalent lipid fraction.}
    \label{figAllLipids}
    }
\end{figure}

\envSubSection{Enrichment as a function of protein--membrane distance}
The enrichment ratio, $\psi_i/\phi_i$, was calculated for the
association of a charged protein with a membrane composed of 69\%
neutral (zwiterionic), 30\% monovalent and 1\% trivalent lipids (in
the absence of a protein). This implies an average charge density
$\SigmaAv = -0.33 e$ per lipid headgroup area. \We{} present the
enrichment arising from a protein that is slightly more charged than
the membrane, $\SigmaProt = - 1.3\SigmaAv$. (These values are typical
to protein--membrane interactions, as will be demonstrated later.)
Figure \ref{figAllLipids} shows the enrichment ratios as a function of
protein--membrane distance.

Far from the membrane ($h \gg \kappa^{-1}$) the protein charge is
screened and its effect on the membrane is weak. As the charged
protein approaches the membrane, oppositely charged lipids move into
the \defBindingDomain{} while the neutral lipids are depleted from it.
The choice of similar charge densities (in absolute value) for the
protein and membrane leads to minor changes in the fractions of the
abundant (neutral and monovalent) lipids, even at distances smaller
than $\kappa^{-1}$.  Notably, the fraction of trivalent lipids changes
by a much larger factor of about 3. This result is a consequence of
the exponential dependence of the enrichment on lipid valence as seen
in \defEquation{eqRelationEquation}.

The nonmonotonic behavior at small distances, shown in
\defFigure{figAllLipids}, is a delicate point that deserves further
discussion. If the membrane charge density had a \emph{fixed} value
$\SigmaDom \neq -\SigmaProt$, then, at a sufficiently short distance,
the mutual attraction between the surfaces would turn into repulsion
(Parsegian and Gingell, 1972). This is caused by the increased
concentration of the salt ions, which are bound to remain in the
confined volume between the protein and membrane in order to
neutralize the system. In \our{} case, however, the system has the
additional freedom to change $\SigmaDom$. As a result, the
electrostatic contribution to the free energy of the protein--membrane
interaction decreases monotonously with decreasing distance, i.e., the
interaction is purely attractive (cf.
\defFigure{figFreeEnergy}). 

{(The effect of charge mobility on the
interaction between two membranes has been studied in detail recently;
Russ et al., 2003.)}~ As long as the two objects are not too close, it
may become favorable to overcharge the membrane and gain attraction
energy.  This is what happens in the system of
\defFigure{figAllLipids} for $\kappa h < 1$.  For example, at $\kappa
h = 0.5$, \we{} find $|\SigmaDom/\SigmaProt|
\simeq 1.17$.  In such a case of overcharging, as the distance is
further reduced, the osmotic pressure of the salt ions at short
distances causes the membrane to decrease its charge density in order
to lower the energetic penalty of further compression
(\defFigure{figAllLipids} in the range $\kappa h \lesssim 0.5$). At
contact ($\kappa h$ = 0) we have $\SigmaDom = -\SigmaProt$, such that
the system is neutral without mobile ions.  Thus, the ability to
redistribute the lipids allows the system to avoid high concentration
of ions in the solution.

\begin{figure}[\envPictureLocation{}]
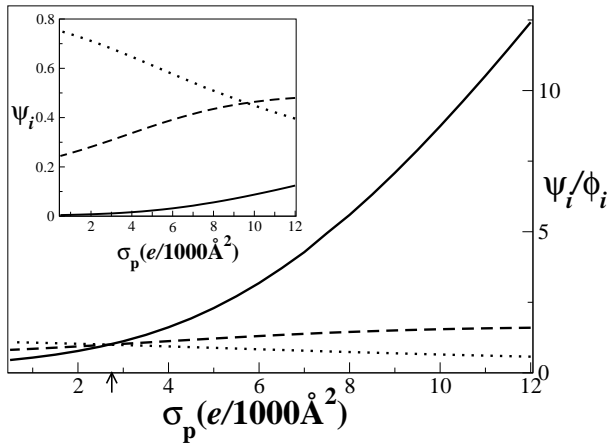

  \center{
    \envIncludeEps{fig3.eps}{3.15in}
    \caption[Protein charge effect]{\envCaptionFont{}Protein charge
      effect. Enrichment ratios ($\psi_i/\phi_i$) of the uncharged
      (dotted), monovalent (dashed) and trivalent (solid) lipids in
      the \defBindingDomain{} of a protein as a function of its charge
      density (in units of elementary charge per $1000$\AA$^2$). The
      inset shows the actual lipid fractions.  The unperturbed
      membrane contains 69\%/30\%/1\% uncharged/monovalent/trivalent
      lipid fractions, corresponding to $\SigmaAv = -4.7e$ per
      1000\AA$^2$ and the protein--membrane distance is $\kappa h =
      0.3$. The enrichment in charged lipids increases with protein
      charge. The arrow indicates the $\SigmaProt$ value for which the
      membrane is unperturbed.}
    \label{figProteinCharge}
    }
\end{figure}

\envSubSection{Effect of protein charge} In
\defFigure{figProteinCharge} \we{} present the enrichment ratios
($\psi_i/\phi_i$) of the different lipid species as a function of the
protein charge density for a given protein--membrane distance and
membrane composition. As expected, when the protein is highly charged,
the \defBindingDomain{} is strongly enriched with oppositely charged
lipids. On the other hand, at low protein charge, depletion of charged
lipids is observed. Remarkably, in both cases, the trivalent species
exhibits a much stronger effect than the monovalent one.  This is
again a consequence of the sensitive dependence of the enrichment
ratio on lipid valence (\defEquation{eqRelationEquation}).  In between
the strongly charged and weakly charged limits, there is a value of
$\SigmaProt$ for which the membrane is unperturbed (see arrow in
\defFigure{figProteinCharge}).  This point does \emph{not} correspond
to $|\SigmaProt| = |\SigmaAv|$, as might have been expected. As
discussed in the previous subsection, for a nonzero protein--membrane
distance the membrane may become overcharged. As a result, this
special point where $\psi_i/\phi_i = 1$ is obtained for $|\SigmaProt|
< |\SigmaAv|$. For example, for the parameters of
\defFigure{figProteinCharge} it occurs at $|\SigmaProt/\SigmaAv|
\simeq 0.57$.

\begin{figure}[\envPictureLocation{}]
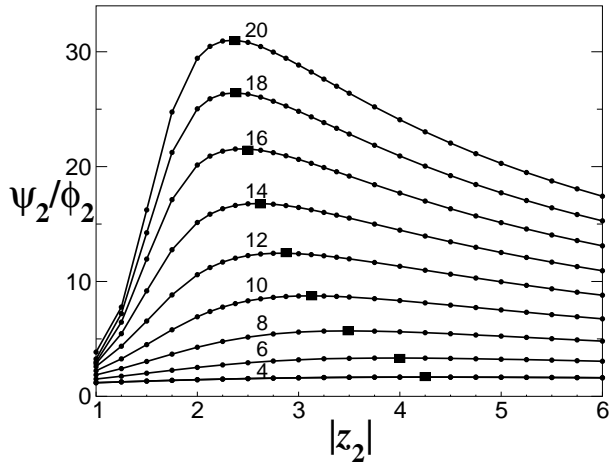

  \center{ \envIncludeEps{fig4.eps}{3.15in}
    \caption[Valence effect]{\envCaptionFont{}Valence effect. The
      enrichment ratio of polyvalent lipids ($\psi_2/\phi_2$) in the
      \defBindingDomain{} of a protein as a function of the lipid
      valence $z_2$.  The different curves correspond to different
      protein charges, the number on each curve indicating the number
      of elementary charges per 1000\AA$^2$. Maximum enrichment is
      marked by a box. The unperturbed membrane composition and
      protein--membrane distance are the same as in
      \defFigure{figProteinCharge}. }
    \label{figOptimalValence}
    }
\end{figure}

\envSubSection{Effect of lipid valence} Equation
\ref{eqRelationEquation} implies that, the higher the valence $|z_2|$
of the lipid, the stronger its enrichment, $\psi_2/\phi_2$, relative
to that of the monovalent species, $\psi_1/\phi_1$. This does
\emph{not} imply that $\psi_2/\phi_2$ \textit{per se} (not relative to
$\psi_1/\phi_1$) increases monotonously with $|z_2|$. In fact, there is
a competition between two opposing effects. The first, which is
entropy-driven, favors charging of the \defBindingDomain{} by
high-valence lipids in order to minimize the perturbation of membrane
composition. On the other hand, from simple stoichiometry, only a
small concentration of a high-valence lipid is needed to attain a
given membrane charge density.  The competition should result in a
maximum of $\psi_2/\phi_2$ at a certain value of valence $z_2 =
z_2^\ast$.  This is confirmed in \defFigure{figOptimalValence}, where
\we{} present $\psi_2/\phi_2$ as a function of $z_2$ for a wide range
of protein charge densities. (To obtain smoother curves \we{}
calculated $\psi_2/\phi_2$ also for artificial, noninteger values of
$z_2$.) For high protein charge, where electrostatic interactions are
strong, the entropy effect is negligible and $|z_2^\ast|$ is small;
for unphysically high protein charge $|z_2^*|$ eventually tends to
$|z_1| = 1$. At low protein charge, entropy dominates and $|z_2^\ast|$
increases. \envThesisRemark{The latter two statements are obtained
  analytically.}

As seen in \defFigure{figOptimalValence}, $|z_2^\ast|$ does not
drastically change with protein charge. For reasonable, physiological
charge densities ($\SigmaProt$ of up to $\sim 10e/1000$\AA$^2$) \we{}
find $|z_2^\ast| \sim$ 3 to 4. Thus, lipid enrichment due to the
adsorption of an oppositely charged protein will be most effective for
a certain lipid valence which, within \our{} idealized model
assumptions, seems to be around ($-3$) to ($-4$).

\envThesisRemark {\input{thesis_completion.tex} }

\envSection{APPLICATION TO PEPTIDE--MEMBRANE INTERACTIONS} In this
section \we{} compare the qualitative results of \our{} model with
recent experimental studies of the lateral sequestration of the
polyvalent lipid PIP$_2$ by adsorbed basic peptides. It should be
borne in mind that \our{} simple model can only provide an approximate
description of such systems.  Treating the interaction region of the
peptide as a large flat surface is a particularly severe
simplification. \We{} return to this and other weaknesses of the model
in the Discussion below.

\We{} focus on two peptides for which there are available experimental
data: \defFAMARCKS{} and a polylysine chain of 13 residues, \defLys{}.
The former corresponds to the basic effector segment of the MARCKS
protein, where five alanine residues were substituted for the original
phenylalanine ones (\defRefGambhir{}; \defRefDiana{}). \We{} avoid
dealing with the MARCKS peptide, since experiments indicate that its
hydrophobic phenylalanine residues pull the peptide into the membrane
such that its backbone penetrates the membrane (Qin and Cafiso, 1996;
Zhang et al., 2003).  This kind of interaction is expected to be
sensitive to specific molecular details, which are not encompassed by
\our{} model.

In order to apply the model, \we{} need an estimate for the area of
the peptide that interacts with the membrane in order to determine the
peptide charge density. In addition, this area is assumed to be
equal to that of the \defBindingDomain{}, $A$ (Fig. 1).
\We{} built extended peptides (MOE software, 2002), similar to the
ones used by Wang et al. (\defRefManuscript{}). \We{} then defined the
effective peptide area as the area of its projected backbone plus an
envelope of width $\kappa^{-1}$ around it. (This somewhat arbitrary
definition will be further examined at the end of this section.)
Dividing the number of charged residues in the peptide by this area,
\we{} got the estimates $\SigmaProt \simeq 13e/2120$\AA$^2$ and
$13e/1060$\AA$^2$ for \defFAMARCKS{} and \defLys{}, respectively. Note
that for these calculated areas and typical peptide--membrane
distances $h$ of a few angstroms, the basic assumption of the model,
$h \ll \sqrt{A}$, is well satisfied.

\We{} used these estimated $\SigmaProt$ values to produce
\defFigure{figAlokSequesterMARCKSLys}, plotting the trivalent lipid
fraction in the \defBindingDomain{} of each peptide as a function of
its distance from the membrane.  In accordance with experiments
(\defRefGambhir{}), \we{} took the unperturbed membrane composition to
be: $\phi_0 = 82\%$ (corresponding to the uncharged zwiterionic PC
lipid), $\phi_1 = 17\%$ (monovalent PS), and $\phi_2 = 1\%$ (trivalent
PIP$_2$). Figure \ref{figAlokSequesterMARCKSLys} shows that the
PIP$_2$ fraction rises to 18\% in the \defLys{} \defBindingDomain{},
whereas only 6\% PIP$_2$ is obtained in the case of \defFAMARCKS{}.
This is caused by the higher charge density of \defLys{}, roughly
double that of \defFAMARCKS{}.  However, \defFAMARCKS{} has twice the
effective area of \defLys{}; thus, if we examine the average
\emph{number} of PIP$_2$ molecules per peptide \defBindingDomain{},
$N_2$, the difference is less significant---while $N_2 \sim 2.7$
for \defLys{} at $\kappa h \lesssim 0.3$, for \defFAMARCKS{} at the
same distance $N_2 \sim 1.8$. (The value of $\kappa h \approx 0.3$
corresponds to $h \approx 3$\AA{} at 100mM salt, which is the
approximate peptide--membrane distance; Ben-Tal et al., 1996; Murray
et al., 2002.  Consequences of this small value will be addressed in
the Discussion below.)
\begin{figure}[\envPictureLocation{}]
  \center{
    \envIncludeEps{fig5.eps}{3.15in}
    \caption[Comparison between lipid rearrangement induced by
    \defFAMARCKS{} and \defLys{}]{\envCaptionFont{}Comparison between
      lipid rearrangement induced by \defFAMARCKS{} and \defLys{}.
      Trivalent lipid fraction in the \defBindingDomain{} of the
      \defFAMARCKS{} (dashed) and \defLys{} (solid) peptides as a
      function of their distance from the membrane. The unperturbed
      membrane is composed of 82\%/17\%/1\%
      uncharged/monovalent/trivalent lipid fractions.  The enrichment
      caused by \defLys{} is much stronger than that achieved by
      \defFAMARCKS{} due to its higher (roughly double) charge
      density.}
    \label{figAlokSequesterMARCKSLys}
    }
\end{figure}

\begin{figure}[\envPictureLocation{}]
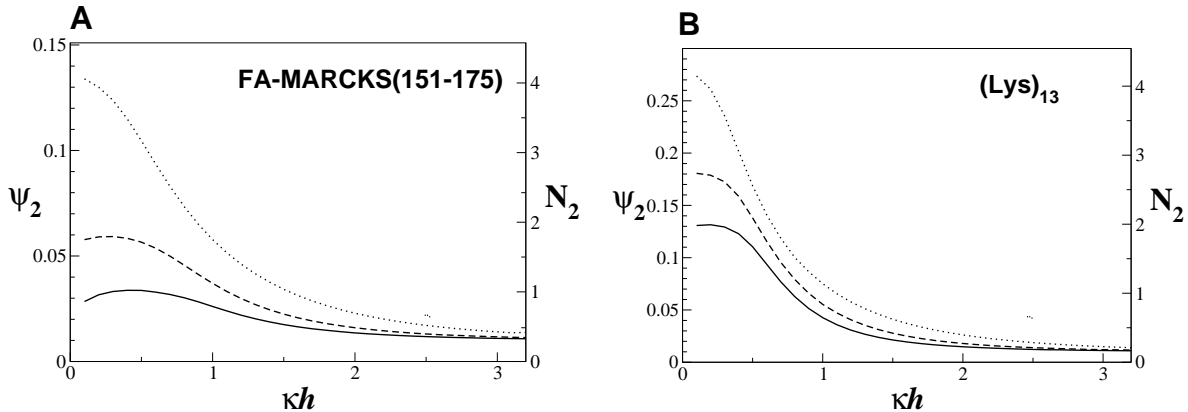

  \centering
  {
    \label{figAlokPSDependence:A}
    \envIncludeEps{fig6a.eps}{3in}
    }
  {
    \label{figAlokPSDependence:B}
    \envIncludeEps{fig6b.eps}{3in}
    }
  \caption[Effect of monovalent lipid concentration]{
    \envCaptionFont{}Effect of monovalent lipid concentration.
    Trivalent lipid fraction (left ordinate) and number (right
    ordinate) in the \defBindingDomain{} of (A) \defFAMARCKS{} and (B)
    \defLys{} peptides as a function of the peptide--membrane
    distance.  Different curves correspond to membranes of different
    lipid compositions: 69\%/30\%/1\% (solid), 72\%/17\%/1\% (dashed),
    and 99\%/0\%/1\% (dotted). The enrichment in the polyvalent lipid
    increases with decreasing monovalent-lipid fraction.}
  \label{figAlokPSDependence} 
\end{figure}

Next, \we{} examined the dependence of the trivalent lipid (PIP$_2$)
fraction in the \defBindingDomain{} on $\phi_1$, the monovalent lipid
fraction in the unperturbed membrane. Figure \ref{figAlokPSDependence}
presents the results obtained for \defFAMARCKS{} and \defLys{} at
various distances. The smaller the value of $\phi_1$, the stronger the
enrichment in trivalent lipids. When there is a little amount of
monovalent lipids, the membrane charge density induced by the
peptide--membrane interaction is attained primarily by the trivalent
species. Therefore, at $\phi_1 = 0$, the enrichment in trivalent lipid
is maximum. At that limit, the number of \defPIP{} molecules per
\defBindingDomain{} can be simply approximated (at short distances) as
the number of charges on the peptide divided by the lipid valence,
$N_2 \approx |A \SigmaProt /(z_2 e)|$. (In practice, however, this
example of the polyvalent lipid being the majority charge is
problematic, as will be presented in the Discussion.) Figure
\ref{figAlokPSDependence} shows that, under physiological conditions
of 30\% monovalent lipid fraction and only 1\% \defPIP{}, both
peptides sequester PIP$_2$---roughly one molecule per \defFAMARCKS{}
peptide and around two molecules per \defLys{} peptide. This result is
in qualitative agreement with experiments (\defRefGambhir{}), where
\defLys{} was found to attract \defPIP{} more strongly than \defFAMARCKS{}.

To emphasize the strong sequestration of the trivalent lipid, consider
the case of a 73\%/17\%/0 PC/PS/PIP$_2$ membrane interacting with
\defLys{}. At a short distance the peptide charge will be neutralized by 
about 13 PS molecules. For a 72\%/17\%/1\% membrane, as shown by the
dashed curve in Fig.\ \ref{figAlokPSDependence}, there are roughly
2--3 sequestered PIP$_2$ lipids per \defLys{}, providing 6--9
elementary charges. Thus, introducing one percent PIP$_2$ into a
membrane of 17\% PS leads to replacement of about one half of the PS
lipids in the \defBindingDomain{} by PIP$_2$. Indication of such an
exchange of PS for PIP$_2$ upon addition of
a small amount of PIP$_2$ was found in a recent experiment
(S.\ McLaughlin, personal communication).

\begin{figure}[\envPictureLocation{}]
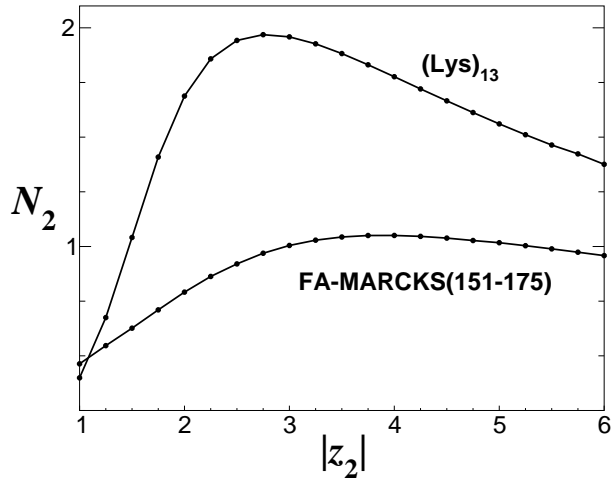

  \center{ \envIncludeEps{fig7.eps}{3.15in}
    \caption[Effect of hypothetical \defPIP{} valence on
    sequestration]{\envCaptionFont{}Effect of hypothetical \defPIP{}
      valence on sequestration. Average number of polyvalent lipids
      per \defBindingDomain{} of \defFAMARCKS{} and \defLys{} as a
      function of \defPIP{} valence. Concentration is a maximum for
      $|z_2| \sim$ 3--4. Both calculations were performed using a
      membrane composition of 69\%/30\%/1\%
      uncharged/monovalent/polyvalent lipid fractions in the
      unperturbed membrane, and a peptide--membrane distance of
      $\kappa h = 0.3$.}
    \label{figPIP2ValMARCKSLys}
}
\end{figure}

Interestingly, the concentration, on average, of about two \defPIP{}
molecules per
\defLys{} would not be possible if \defPIP{} were of much different
valence. Figure \ref{figPIP2ValMARCKSLys} shows the average number of
\defPIP{} molecules per \defBindingDomain{} of both peptides as a
function of the \defPIP{} hypothetical valence. Similar to the results
shown in \defFigure{figOptimalValence}, we find a nonmonotonic
behavior as a function of valence with a maximum at $|z_2| \sim$ 3 to
4.  It is stressed again that, in view of \ourthis{} simplified model,
one should pay more attention to the existence of a competition
mechanism, leading to an optimum valence, than to the exact value
obtained for that valence.

\begin{figure}[\envPictureLocation{}]
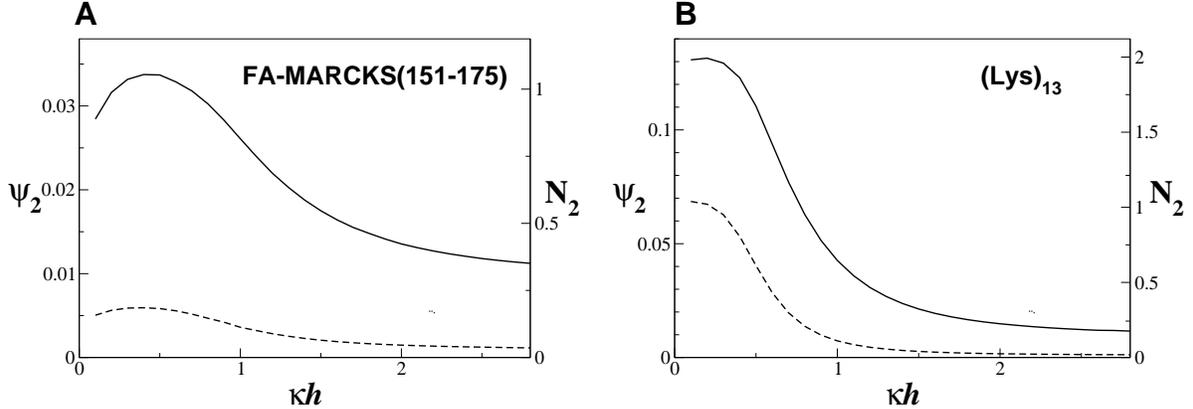

  \centering
  {
    \envIncludeEps{fig8a.eps}{3in}
    }
  {
    \envIncludeEps{fig8b.eps}{3in}
    }
  \caption[Effect of polyvalent lipid fraction in the bare
  membrane]{ \envCaptionFont{}Effect of $\phi_2$ on enrichment.
    Trivalent lipid fraction (left ordinate) and average number (right
    ordinate) in the \defBindingDomain{} of (A) \defFAMARCKS{} and (B)
    \defLys{} peptides as a function of the peptide--membrane
    distance.  The two curves correspond to membrane compositions of
    69\%/30\%/1\% PC/PS/\defPIP{} (solid) and 69.9\%/30\%/0.1\%
    PC/PS/\defPIP{} (dashed).}
  \label{figAlokPIPDependence} 
\end{figure}

In \defFigure{figAlokPIPDependence} \we{} show the dependence of the
trivalent lipid fraction in the \defBindingDomain{} on its value in
the unperturbed membrane. As expected, $\psi_2$ increases with
$\phi_2$. This calculation shows that one needs $\phi_2 \gtrsim 1\%$
in order to get an average stoichiometry of 1:1 between \defFAMARCKS{}
and \defPIP{}. As we have seen above, \defLys{} sequesters \defPIP{}
more effectively. Hence, as demonstrated in
\defFigure{figAlokPIPDependence}, a $\phi_2$ value of only $0.1\%$ is
sufficient to obtain a 1:1 \defLys{}:\defPIP{} ratio. That is,
\defLys{} can sequester an appreciable amount of \defPIP{} even when
the membrane contains 300-fold more PS lipids than \defPIP{}. This
value is in very good agreement, perhaps fortuitously, with the
measurements of Gambhir et al. (\defRefManuscript{}).

Finally, \we{} examined the dependence of the \defPIP{} enrichment on
the ionic strength, i.e., the concentration of mobile salt ions in the
solution, $n_0$. This parameter enters into the model through the Debye
screening length $\kappa^{-1}$, which both scales the distance $h$ and
affects the amplitude of the electrostatic potential (see, e.g.,
\defEquation{eqParsegianPsi} in Appendix I). Figure
\ref{figSaltDependence} shows that $\psi_2$ decreases with ionic
strength. This is a result of the increased screening of the
electrostatic attraction between the membrane and protein. The changes
are not dramatic up to quite high $n_0$ values. The reason is the very
close proximity of the two objects (3\AA) for which $\kappa h < 1$ in
the entire $n_0$ range examined.

\begin{figure}[\envPictureLocation{}]
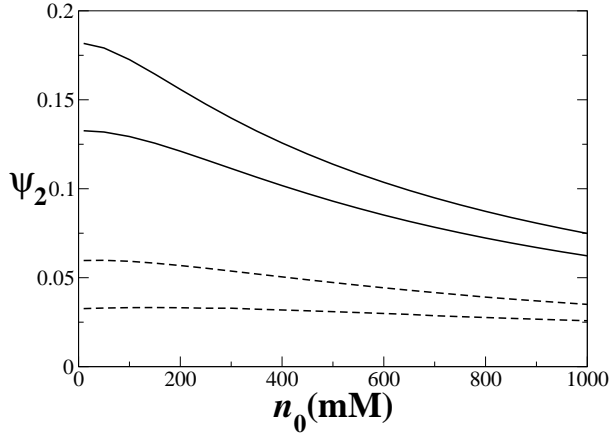

  \center{
    \envIncludeEps{fig9.eps}{3.15in}
    \caption[Ionic strength effect]{\envCaptionFont{}Ionic strength
      effect. Trivalent lipid fraction in the \defBindingDomain{} of
      \defFAMARCKS{} (dashed) and poly-lysine (solid) peptides as a
      function of salt concentration. The two curves for each peptide
      were calculated using membrane compositions of 69\%/30\%/1\%
      (upper), and 82\%/17\%/1\% (lower).  A peptide--membrane
      distance of $h = 3$\AA{} was used. }
    \label{figSaltDependence}
    }
\end{figure}

\envSubSection{Effect of approximated peptide size} As discussed
above, \we{} define the effective area of the peptide as the area of
its projected backbone plus an envelope of width $\kappa^{-1}$ around
it.  This definition, however, is somewhat arbitrary.  We therefore
examine the effect of relaxing the effective-area definition on the
values obtained for $N_2$.

\begin{figure}[\envPictureLocation{}]
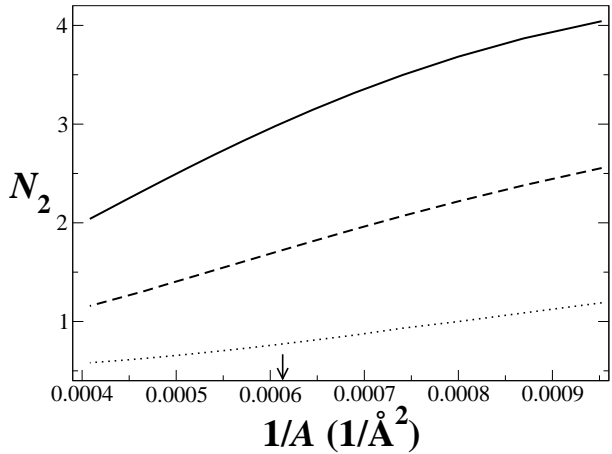

  \center{
    \envIncludeEps{fig10.eps}{3.15in}
    \caption[Number of polyvalent lipid dependence on measured area
    of peptide]{\envCaptionFont{}Dependence of polyvalent lipid number
      on estimated peptide area. Average number of trivalent lipids
      ($N_2$) as a function of the inverse area of the
      \defBindingDomain{} $1/A$.  $A$ is also taken as the effective
      peptide area.  The arrow marks the $1/A$ value according to the
      definition used throughout this study.  Results were obtained
      for a 10 (dotted), 15 (dashed) and 20 (solid) charged residues
      in a 20-amino-acid-long peptide interacting with a 69\%/30\%/1\%
      uncharged/monovalent/trivalent membrane at a distance of $\kappa
      h = 0.3$. }
    \label{figLinearResponse}
    }
\end{figure}

\begin{figure}[\envPictureLocation{}]
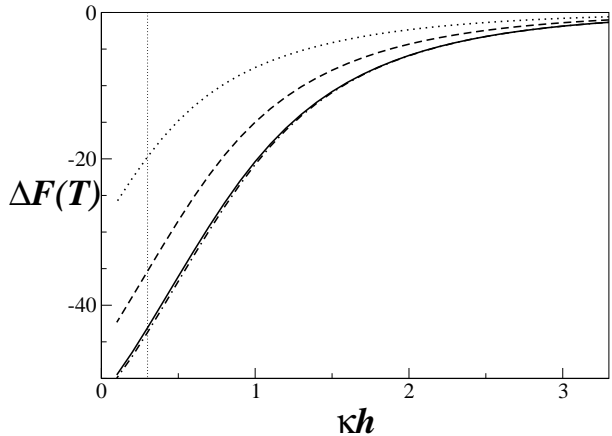

  \center{
    \envIncludeEps{fig11.eps}{3.15in}
    \caption[Contribution of electrostatics and entropy to the
    free energy of peptide--membrane interaction]{\envCaptionFont{}
      Contribution from electrostatics and entropy within the smeared
      charge model to the free energy of peptide--membrane
      interaction. Free energy in units of $T$ as a function of the
      peptide--membrane distance.  The curves correspond to
      \defFAMARCKS{} and membrane compositions of 69\%/30\%/1\%
      (solid), 77\%/18\%/5\% (dash-dotted), 72\%/17\%/1\% (dashed),
      and 90\%/10\%/0\% (dotted). The vertical dotted line indicates a
      distance of $\kappa h = 0.3$. The presented contribution to the
      free energy decreases monotonously with distance. The solid and
      dash-dotted curves correspond to membranes of very different
      lipid compositions but the same charge density ($0.33e/a$).  }
    \label{figFreeEnergy}
    }
\end{figure}

Figure \ref{figLinearResponse} shows the average number of trivalent
lipids $N_2$ per \defBindingDomain{} of a 20-amino-acid-long peptide
for a range of $1/A$ values extending to $\pm 50\%$ of \our{} original
estimate. In this range the alteration in $N_2$ is limited to $\pm1$
lipid molecules.

\envSubSection{Interaction free energy} 
From the Poisson-Boltzmann
theory, as applied to \ourthis{} model, \envArticleRemark{we can
  calculate} the contribution to the free energy of peptide--membrane
association coming from electrostatics and entropy\envThesisRemark{
  can be calculated}. The derivation is given in Appendix II.  The
results for the case of \defFAMARCKS{} interacting with a membrane of
different compositions are presented in \defFigure{figFreeEnergy}.
Some of these compositions have also been studied by Gambhir et al.
(\defRefManuscript{}) and Wang et al. (\defRefManuscript{}). It should
be recalled that \ourthis{} model does not take into account repulsive
interactions, such as the Born desolvation effect (see 
\defRefDiana{}). Lipid demixing effects are included in the
model, so the local membrane charge density may change due to the
peptide.
As a result, the electrostatic free energy in \ourthis{} model 
decreases monotonously as \defFAMARCKS{} approaches the membrane. To
get the \emph{total} free energy of association, one ought to add a
repulsive interaction at short distances (not included in \ourthis{}
model), that would yield a free-energy minimum at a distance of a few
angstroms (see, e.g., \defRefDiana{}).  Thus, the free-energy values
presented in \defFigure{figFreeEnergy} are probably more negative than
the actual binding free energy. 
More detailed models provide values for the total binding free energy
that are roughly one half the contribution presented here
(\defRefDiana{}).

As expected, electrostatic attraction between the peptide and the
membrane is proportional to the charge density of the membrane
(\defFigure{figFreeEnergy}). The interaction free-energy is not
sensitive to the specific lipid composition of the membrane, but
rather to its average charge density. This result stems from the minor
contribution of mixing entropy to the free energy (fourth term in
\defEquation{eqFreeChargesReduced}) and the approximate behavior of
the membrane in our model as a surface of constant electric potential,
which is determined by the average charge density $\overline\sigma$
(May et al., 2000). The calculated contribution to the free energy,
therefore, is mainly the work required to bring a charged object
(protein) into such a potential, regardless of the lipid
redistribution. This may also be the reason why FDPB calculations
(e.g. Ben-Tal et al., 1996; Murray et al., 2002), although ignoring
lipid redistribution, often give a good estimate of the binding free
energy.

\envSection{DISCUSSION} \envSubSection{Model evaluation} A 
simplified theoretical model for the effect of an adsorbed charged
protein on the distribution of phospholipids in the membrane has been
presented. One of the key theoretical results is the relative
enrichment equation,
\defEquation{eqRelationEquation}, relating the concentrations of
various lipid species in the \defBindingDomain{}.  It should be
emphasized that this result is almost model-independent.  Within a
two-block picture, it should hold for any protein--membrane system
governed by electrostatics, regardless of particular details of the
protein, the strength of the electrostatic interactions, and the
validity of a mean-field (PB) assumption. The main physical effect
described by this relation is the sensitivity of lipid enrichment to 
valence, i.e.,
the increased concentration of polyvalent lipids compared to that of
monovalent ones in the adsorption domain.

We have demonstrated how simple Poisson-Boltzmann calculations can be
added to \defEquation{eqRelationEquation} to obtain further details of
the membranal interaction region. In most of the biologically relevant
cases, including the peptides discussed here, the peptide--membrane
distance $h$ is of the order of a few angstroms only (Ben-Tal et al.,
1996; 1997; Murray et al., 2001; 2002). In the context of \our{}
model, the elaborate Poisson-Boltzmann theory is unnecessary for such
a thin water layer between the two charged objects. Nevertheless, it
should be recalled that, at any rate, \our{} calculations converge to
the correct membrane charge density at small $h$ 
{due to charge neutrality,} 
$\SigmaDom = -\SigmaProt$ (cf.\ \defFigure{figCompareMethods}).

The model is focused on the qualitative behavior of electrostatically
dominated protein--membrane systems. \Our{} description of the
protein--membrane interaction is evidently crude, as it omits the
molecular details of this complex system; see the atomistic
description used in the \envArticleRemark{companion} report of Wang et
al.  (\defRefManuscript{}). Thus, effects related to charge
discreteness and three-dimensional structure are ignored. Such
effects, for example, may play an important role in the
membrane adsorption of the MARCKS protein, 

{whose backbone is believed
to penetrate the membrane (Zhang et al., 2003) and whose charges are
not evenly spaced.}  
Furthermore, \we{} considered only electrostatic
and entropy effects while neglecting other interactions (e.g.,
desolvation). The model is not intended to reproduce such data as the
total binding free energy or the optimum protein--membrane distance
(e.g., Ben-Tal et al., 1996; Arbuzova et al., 2000).  Similarly, the
values that have been obtained for the various lipid concentrations
should be regarded as rough estimates.

Treating a peptide as a large, flat, uniformly charged surface is
probably the gravest simplification of the model, deserving further
discussion. The assumption of flatness may be applicable to proteins
interacting with the membrane through such effector regions as those
described in Wang et al. (\defRefManuscript{}) and Gambhir et al.
(\defRefManuscript{}). The assumption regarding the lateral extent of
the peptide is reasonable in cases where the peptide--membrane
distance $h$ is much smaller than both lateral dimensions. For
example, in the case of the \defFAMARCKS{} discussed above, \we{}
estimated the peptide surface area facing the membrane as a rectangle
of dimensions 106\AA $\times$ 20\AA. The smaller lateral dimension is
still significantly larger than the typical value of $h \sim 3$\AA.

{In addition, the fact that this smaller lateral dimension ($\sim
20$\AA) is comparable to the screening length may lead, in principle,
to significant finite-size effects. Yet, as already noted above,
because of the small value of $h$ in the relevant systems, the charge
density in the adsorption domain should become insensitive to these
details and be determined, to a good approximation, merely by charge
neutrality.}

The small value of $h$, on the other hand, raises a difficulty with
respect to the smeared-charge simplification. Since $h$ is similar to
or smaller than the typical distance between charged groups on the
peptide, neglecting charge discreteness is clearly questionable.
Spreading the charges of the lipid headgroups evenly over the membrane
is problematic as well.  

{Nonetheless, several recent experiments
might help us indicate the limits of validity of models based on
smeared electrostatics, such as ours. The affinity of \defFAMARCKS{}
to PC:PS membranes was found to depend linearly on PS fraction (Murray
et al., 1999). Adding a small amount of \defPIP{} to such membranes,
having a significant PS fraction, did not change the binding
significantly (S. McLaughlin, personal communication).  These two
observations are in line with simple electrostatic considerations.
However, the binding affinity of the same peptide to a membrane of
relatively low charge density, composed of 99:1 PC:\defPIP{}, is
surprisingly large --- similar to that of a 5:1 PC:PS membrane ---
even though the membrane charge densities in the two cases differ by a
factor of about 5 (Wang et al., 2002). Clearly, the latter
observation cannot be accounted for by smeared electrostatics.  We
therefore believe that our model gives reliable results regarding
lipid redistribution only in cases where the membrane has a high
``background'' charge density, i.e., a large fraction of monovalent
lipid, which is the biologically relevant case. For such membranes,
as discussed in the previous section, the model predictions are
in qualitative agreement with available experiments.}

Apart from these assumptions, the model contains another implicit
simplification, namely, that all the electric field lines are
contained within the aqueous spacing between the membrane and protein.
This commonly used assumption is strictly correct in the limit where
the objects are either infinitely thick or of a vanishing dielectric
constant. As \envArticleRemark{will be reported
  elsewhere}\envThesisRemark{have been shown in section 4}, \we{} find
that this approximation is, in fact, still good for objects of
$\epsilon = 2$ and thickness as small as one third the Debye length.

On the positive side, the simplified model presented here provides
new insights into lipid redistribution caused by protein adsorption.
Although most of the results presented in this work were obtained
using a numerical solution of the nonlinear Poisson-Boltzmann
equation, \we{} have demonstrated how one can get the same qualitative
results using a much simpler calculation, the SLPB method, involving
merely a set of polynomial equations (see Appendix I). Such a scheme
may serve as a better starting point for more detailed numerical
calculations, e.g., FDPB (Honig et al., 1993). In most of the current
FDPB calculations, a predefined membrane composition identical to that
of the bare membrane is used (Ben-Tal et al., 1996; 1997; Murray et
al., 1999;
\defRefDiana{}).  This arbitrary description of the
\defBindingDomain{} may be improved if one uses a preliminary analysis
of the type presented here to produce an approximate lipid
configuration.

Future extensions of this work may include phenomena such as elastic
deformation of the membrane (Dan et al., 1993; May, 2000), adsorption
of multiple proteins (May et al., 2000; May et al., 2002), nonuniform
charge density in the \defBindingDomain{} (May et al., 2000), and acid
dissociation at different pH values (Fleck et al., 2002).

\envSubSection{Biological implications} 
ENTH, FYVE, PX and other membrane-association domains use
predefined stereochemistry to recognize poly-phosphoinositides
(McLaughlin et al., 2002; Lemmon, 2003). These domains, which are
commonly found in proteins involved in intracellular signaling, bind
tightly to the poly-phosphoinositides, often via ion pairs, anchoring
the protein firmly and 
irreversibly to membrane surfaces. The binding specificity is
reflected in the evolutionary conservation across the homologous
domains comprising the family; usually the amino acid residues that
mediate the poly-phosphoinositides binding are strictly conserved and
can often serve as sequence signatures to recognize these domains
using sequence analysis tools.

Here \we{} dealt with a much less specific, and often reversible, mode
of membrane recognition via a cluster of basic residues on the
membrane-facing region of the protein. These residues interact
electrostatically with acidic lipids in the \defBindingDomain{} on the
bilayer surface. 
\Our{} model showed that in such cases membrane association induces a
preference for polyvalent lipids such as \defPIP{} to sequester in the
\defBindingDomain{}. The number of sequestered polyvalent lipids may
be regulated by the charge density and size of the
membrane-interaction region on the protein. This result supports the
suggestion, advocated in the \envArticleRemark{accompanying papers
  (\defRefGambhir{}; \defRefDiana{}),} \envThesisRemark{papers of
  Gambhir et al.( \defRefGambhir{}); Wang et al.
  (\defRefManuscript{})} that membrane-associated proteins such as
adducin (Matsuoka et al., 2000), DAKAP200 (Rossi et al., 1999), GAP43
(Laux et al., 2000), MARCKS (Wang et al., 2002), and MacMARCKS
(Blackshear, 1993), which contain a cluster of basic residues, may
create a reservoir of \defPIP{} molecules in their \defBindingDomain{}
(McLaughlin et al., 2002). It may further imply that the
membrane-interaction region of such proteins does not have to be
strictly conserved evolutionarily; it should only preserve a specific
charge density, as in C2 domains (Murray and Honig, 2002).  This
speculation, naturally, needs to be checked in future studies.

\We{} found that, when all other parameters are held fixed, there is
an optimum value of lipid valence that yields maximum enrichment in
the \defBindingDomain{}
(\defFigures{figOptimalValence}{figPIP2ValMARCKSLys}). This value,
resulting from a competition between entropy and stoichiometry,
is found to be at reasonable valence values (e.g., $z_2^\ast = -3$ to $-4$
in the examples above). Thus, if in a certain biological scenario
there is a need to increase the local concentration of a phospholipid
by electrostatic interactions, a polyvalent lipid of valence larger than
1 but not too large (say, $\sim -3$) would be advantageous. The
valence of \defPIP{}, considered to be between $-3$ to $-5$ (Toner et
al., 1988; McLaughlin et al., 2002; Rauch et al., 2002; Wang et al.,
2002) appears to be in line with this criterion.

Assuming a certain valence for \defPIP{} and a certain distance for
peptide association with the membrane, \ourthis{} model enables the
derivation of a set of approximate rules relating the number of basic
residues on the peptide to the average number of sequestered \defPIP{}
molecules per peptide. For example, assuming a trivalent \defPIP{} and
association distance of about 3\AA{}, under physiological conditions,
each segment of seven consecutive lysine residues of an adsorbed
poly-lysine peptide such as (Lys)$_{13}$
(\defFigure{figAlokPSDependence}B) would sequester on the average
approximately one
\defPIP{} molecule when the membrane composition is $69\%/30\%/1\%$
uncharged/monovalent/trivalent, and roughly one and a half \defPIP{}
molecules when the membrane composition is $82\%/17\%/1\%$. These
results are in good agreement with the detailed calculations reported
in \envArticleRemark{the companion report of} Wang et al. (cf. Table
2B in \defRefDiana{}).

\envMakeAcknowledgments{}

\setcounter{secnumdepth}{0}
\appendix
\envSection{APPENDIX I: METHODS FOR CALCULATING LIPID CONCENTRATIONS
    \envThesisRemark{\newline Derivation of
    concentration values}}
\label{secAppNLPB}

In this section \we{} present the method for calculating the values of
$\psi_i$. We present three levels of approximation and compare their
results.

In thermodynamic equilibrium the electrochemical potentials of each
lipid species in the protein-free membrane and in the protein
adsorption domain should be equal, $\defMuBare = \defMuDom$. Within a
mean-field approximation, this condition can be written as
\begin{equation}
  \label{eqCompareMu}
  z_i e \Psi^{(0)}(0) + T \ln{\frac{\phi_i}{\phi_0}} = 
  z_i e \Psi(0) + T \ln{\frac{\psi_i}{\psi_0}} 
\end{equation}
where $\Psi^{(0)}(0)$ is the mean electrostatic potential at the bare
membrane, and $\Psi(0)$ its value at the \defBindingDomain{}. 
Equation \ref{eqCompareMu} is actually a set of ($k - 1$) equations
for every species $i \neq 0$.  The incompressibility condition,
$\sum_i\psi_i = 1$, closes a set of $k$ equations for the $k$ unknown
$\psi_i$. To solve these equations \we{} need the surface potentials
$\Psi^{(0)}(0)$ and $\Psi(0)$. \We{} derive them using the
Poisson-Boltzmann theory.  Note that $\Psi(0)$ depends on the
variables $\psi_i$ that determine the charge density of the
\defBindingDomain{}, $\SigmaDom$.

\envSubSection{Nonlinear Poisson-Boltzmann} To find the surface
potentials \we{} need to solve the Poisson-Boltzmann equation,
\begin{equation}
  \label{eqPB}
  \Dscd{y}{z} = \kappa^2\sinh{y}
\end{equation}
with the appropriate boundary conditions. In \defEquation{eqPB} $y$ is
the local dimensionless potential at a distance $z$ from the membrane,
$y(z) \equiv e\Psi(z)/T$. \envThesisRemark{Note that this equation
  holds only for salt comprised of monovalent ions.}

For the bare membrane the boundary conditions are
\begin{equation}
  \left. \Dfst{y^{(0)}}{z}\right|_{z=0} = -\frac{4 \pi e \SigmaAv}{\epsilon T}
\end{equation}
and vanishing of the field at $z \rightarrow \infty$.  This problem,
of a single charged plate in an electrolyte, is analytically solvable
in closed form (Andelman, 1995). The result is
\begin{equation}
  \label{eqAndelmanPsi}
  y^{(0)}(0) = -4\tanh^{-1}\gamma
\end{equation}
where $\gamma$ is the positive root of the quadratic equation
$\gamma^2 + \gamma \kappa \epsilon T/(\pi e |\SigmaAv|) = 1$. (\We{}
have assumed the membrane to be negatively charged.)

For the protein--membrane system, the boundary conditions are
\begin{equation}
  \label{eqBoundCond}
  \left. \Dfst{y}{z}\right|_{z=0} = -\frac{4 \pi e \SigmaDom}{\epsilon T}  \;\;\; , \;\;\;
  \left. \Dfst{y}{z}\right|_{z=h} = \frac{4 \pi e \SigmaProt}{\epsilon T}
\end{equation}
Integrating the Poisson-Boltzmann equation (\defEquation{eqPB}) once
while applying both boundary conditions, \we{} get
\begin{eqnarray}
  \label{eqPBPartialSolution}
  &&\Dfst{y}{z} = \kappa\sqrt{2\cosh{y} + C}  \;\;, \;\;
  C = \left(\frac{4 \pi e \SigmaDom}{\epsilon \kappa T}\right)^2 - 2\cosh{y(0)}\\
  \label{eqNLEq1}
  &&\cosh{y(0)}-\cosh{y(h)} = 2\left(\frac{2\pi e}{\epsilon \kappa T}\right)^2(\SigmaDom^2 - \SigmaProt^2)
\end{eqnarray}
\We{} then integrate \defEquation{eqPBPartialSolution} to get
\begin{equation}
  \label{eqNLEq2}
  h = \int_0^h{dz} = \frac{1}{\kappa \sqrt{2}}\int_{y(0)}^{y(h)}{\frac{dy}{\sqrt{\cosh{y} + C/2}}}
\end{equation}
Equations \ref{eqNLEq1} and \ref{eqNLEq2} are solved numerically for
the two unknowns $y(0)$ and $y(h)$, the surface potentials of the
membrane and protein. Finally, the values of $y(0)$ and $y^{(0)}(0)$
(\defEquation{eqAndelmanPsi}) are used in \defEquation{eqCompareMu} to
calculate the values of $\psi_i$.

\envSubSection{Linear Poisson-Boltzmann}
\label{secAppLPB}
The LPB approximation holds when the electrostatic interactions are
weak compared to $T$, $y \ll 1$. (In fact, the systems relevant to our
study are far from this limit; cf.\ \defFigure{figFreeEnergy}.)  In
this limit \we{} can linearize the Poisson-Boltzmann equation
(\defEquation{eqPB}),
\begin{equation}
  \label{eqLPB}
  \Dscd{y}{z} = \kappa^2y
\end{equation}
The surface potential of the unperturbed membrane then takes the form
(Evans and Wennerstrom, 1994),
\begin{equation}
  \label{eqEvansPsi}
  y^{(0)}(0) = \frac{4 \pi e \SigmaAv}{\epsilon \kappa T}
\end{equation}

For the protein--membrane system \we{} integrate \defEquation{eqLPB}
twice while applying the boundary conditions (\defEquation{eqBoundCond})
to obtain (Parsegian and Gingell, 1972)
\begin{equation}
  \label{eqParsegianPsi}
  y(0) = \frac{4 \pi e} {\epsilon \kappa T \sinh{(\kappa
      h)}} [\SigmaProt + \SigmaDom \cosh{(\kappa h)}]
\end{equation}
The potential values of \defEquations{eqEvansPsi}{eqParsegianPsi} are
then used in \defEquation{eqCompareMu} to calculate the $\psi_i$
values.

\envSubSection{Simplified linear Poisson-Boltzmann}
\label{secAppSLPB}
In this further approximation the entropy is neglected in $\defMuBare$
and $\defMuDom$, and we are left with a uniform surface potential
$y(0)=y^{(0)}(0)$. 

\begin{figure}[\envPictureLocation{}]
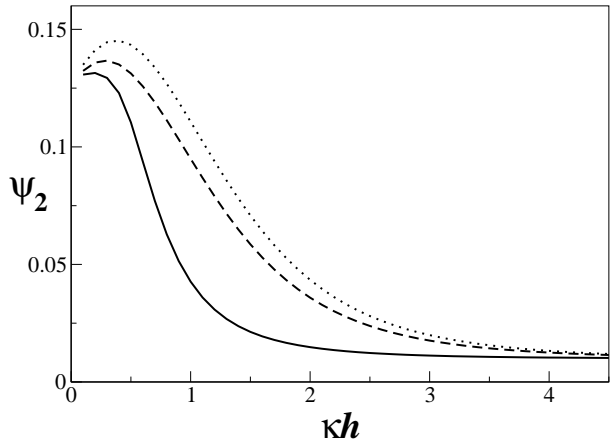

  \center{
    \envIncludeEps{fig12.eps}{3.15in}
    \caption[Comparison of the different computational
    methods]{\envCaptionFont{}Comparison of the different
      computational methods. Trivalent lipid fraction in the
      \defBindingDomain{} of a highly charged protein ($\SigmaProt =
      13e/1000$\AA) as a function of the protein--membrane distance.
      Results were obtained using three methods: nonlinear
      Poisson-Boltzmann (solid curve), linear Poisson-Boltzmann
      (dashed curve), and the simplified linear Poisson-Boltzmann
      (dotted curve). At distances $\kappa h < 0.3$, where the
      membrane approaches charge matching, the three methods differ by
      less than 10\%.  The unperturbed membrane composition is the
      same as in \defFigure{figProteinCharge}.}
    \label{figCompareMethods}
}
\end{figure}

{Note that the entropy contribution is not
necessarily small compared to $T$. The strong enrichment in trivalent
lipid, as demonstrated in Figs.\ \ref{figProteinCharge} and
\ref{figCompareMethods}, entails an entropy penalty of a few $T$. Yet,
it is still much smaller than the electrostatic contribution, which
amounts to tens of $T$ (cf.\ \defFigure{figFreeEnergy}).}

Within the LPB approximation \we{} can equate expressions
\ref{eqEvansPsi} and \ref{eqParsegianPsi} to obtain the charge density
in the \defBindingDomain{} in closed form as given in
\defEquation{eqAnalyticSigma} (May et al. 2000).  Substituting
$\SigmaDom$ in \defEquation{eqRelationSigma} yields a set of $k$
equations, Eqs.  \ref{eqRelationSigma}, \ref{eqSumEq1}, and
\ref{eqRelationEquation}, which are easily solved for the $k$ lipid
fractions $\psi_i$.

\envSubSection{Comparison of the methods} Here
\we{} compare the NLPB, LPB and SLPB calculation methods for a
specific example. In the limit of weak electrostatic interactions
compared to the thermal energy $T$ the LPB method should coincide
with the NLPB one. To highlight the difference between the methods
\we{} therefore chose as an example a highly charged protein ($\SigmaProt =
13e/1000$\AA$^2$). The trivalent lipid fractions in the
\defBindingDomain{}, as calculated using the three methods, are
plotted in \defFigure{figCompareMethods}.  For both long and short
distances all three curves match. In the long distance limit ($\kappa
h \gg 1$), the protein--membrane interaction is weak and the
linearization of the Poisson-Boltzmann equation is valid. At very
short distances the membrane, as described by all three methods, is
forced by charge neutrality to match the charge density of the protein
(in opposite sign).  In the intermediate range the methods differ, yet
because of the two constraints at $\kappa h \ll 1$ and $\kappa h \gg
1$ the differences are mild.  It is noteworthy that at distances
$\kappa h \lesssim 0.3$ (typical to protein--membrane adsorption), the
calculated $\psi_i$ values obtained using the three methods differ by
less than 10\%, and the difference in magnitude between the charge
densities of the protein and membrane are also less than 10\%.

\envSection{APPENDIX II: FREE ENERGY CALCULATION
\envThesisRemark{\newline Derivation of
    free energy}}

In this appendix \we{} calculate the free energy of
protein--membrane interaction as arising from \envArticleRemark{our}
\envThesisRemark{the presented} model. It should be recalled that 
what is presented here is not the total
binding free energy but only the contributions due to simple
electrostatics and entropy. In particular, since charges are allowed
to redistribute in the membrane, the interaction is purely attractive
and the free energy decreases monotonously with distance, reaching its
minimum at $h=0$. This should be contrasted with the FDPB calculations
(e.g., Ben-Tal et al., 1996; \defRefDiana{}) which take desolvation
into account and yield a free-energy minimum at a finite distance.

The general form of the free energy per unit area within a mean-field
approximation is
\begin{eqnarray}
  \lefteqn{\defTotalFDom(h) = \int_0^h\left[-\frac{\epsilon}{8\pi}
    \left(\Dfst{\Psi}{z}\right)^2 + e(n_+-n_-)\Psi \right]dz} \nonumber \\
&& \mbox{}+  \int_0^h [T\left(n_+\ln{n_+}-n_++n_-\ln{n_-}-n_--2n_0\ln{n_0}
+2n_0\right) \nonumber \\
&& -\mu_+ (n_+-n_0) - \mu_- (n_--n_0)] dz \nonumber \\
&& \mbox{} +\frac{T}{a}\sum_i \psi_i \ln{\psi_i}
-\frac{1}{a}\sum_{i>0} \mu_i \psi_i + \SigmaProt  \Psi(h) + \SigmaDom \Psi(0) 
\label{eqTotalGrand}
\end{eqnarray}
The first term in \defEquation{eqTotalGrand} is the energy associated
with the mean electric field. The second term corresponds to the
interaction of the mobile ions with the field, where $n_+$ and $n_-$
are the local concentrations of the monovalent positive and negative
ions.  The second integral accounts for the ideal entropy of mixing of
the mobile ions and their contact with ion reservoirs
having chemical potentials $\mu_+$ and $\mu_-$. The last line in
expression \ref{eqTotalGrand} corresponds to the surface energy of the
membrane and protein, including mixing entropy of the lipids and
electrostatic interactions.

Setting the variations of $F$ with respect to $n_+$, $n_-$, and $\Psi$
to zero, one properly recovers the Boltzmann relations for the mobile
ions, the Poisson equation, and, hence, also the Poisson-Boltzmann
equation.  Using these results along with
\defEquation{eqPBPartialSolution}, and changing the integration
variable from $z$ to $y$, we obtain the following simplified
expression for the free energy:
\begin{equation}
  \defTotalFDom(h) = \frac{2Tn_0}{\kappa}\int_{y(0)}^{y(h)}\frac{
    1 - 2\cosh y -C/2}{\sqrt{2\cosh y + C}}dy +\frac{T\SigmaProt}{e}
  y(h)+\frac{T\SigmaDom}{e}y(0)+ \frac{T}{a}\sum_i \psi_i
  \ln{\psi_i}- \frac{1}{a}\sum_{i>0} \mu_i \psi_i 
  \label{eqFreeChargesReduced}
\end{equation}
where the constant $C$ was defined in
\defEquation{eqPBPartialSolution}.  We can now use the values obtained
for $y(0)$, $y(h)$, $\mu_i$, and $\psi_i$, as described in appendix I,
to calculate the free-energy of interaction $\defTotalFDom(h)$.

\envThesisRemark{\input{thesis_appendix.tex}}

\clearpage
\addcontentsline{toc}{section}{REFERENCES}


\begin{thebibliography}{99}

 
\bibitem{Andelman94}{Andelman, D., M. M. Kozlov, and W. Helfrich.
    1994.}  {Phase transitions between vesicles and micelles driven
    by competing curvatures.  \textit{Europhys. Lett.}  25:231-236.}
  
\bibitem{Andelman95}{Andelman, D. 1995.}  {Electrostatic properties
    of membrane: the Poisson Boltzmann theory. \textit{In} Structure
    and Dynamics of Membranes, 2nd ed, Vol 1B.  R. Lipowsky and E.
    Sackmann, editors. Elsevier, Amsterdam. 603-642.}
  
  
  
\bibitem{Arbuzova00}{Arbuzova, A., L. Wang, J. Wang, G.
    Hangyas-Mihalyne, D. Murray, B. Honig, and S. McLaughlin. 2000.}
  {Membrane binding of peptides containing both basic and aromatic
    residues. Experimental studies with peptides corresponding to the
    scaffolding region of caveolin and the effector region of MARCKS.
    \textit{ Biochem.} 39:10330-9.}

  \envThesisRemark{\bibitem{Bell77}{Bell, G. M., and G. C. Peterson.
      1977.}{Forces between dissimilar colloidal plates for various
      surface conditions .1. general method. \textit{J. Colloid.
        Interf. Sci.} 60:376-385.}}
  
\bibitem{BenTal96}{Ben-Tal, N., B. Honig, R. M. Peitzsch, G. Denisov,
    and S. McLaughlin. 1996.}  {Binding of small basic peptides to
    membranes containing acidic lipids: theoretical models and
    experimental results.  \textit{Biophys. J.} 71:561-575.}
  
\bibitem{BenTal97}{Ben-Tal, N., B. Honig, C. Miller, and S.
    McLaughlin.  1997.}  {Electrostatic binding of proteins to
    membranes.  Theoretical predictions and experimental results with
    charybdotoxin and phospholipid vesicles.  \textit{Biophys. J.}
    73:1717-1727. }
  
\bibitem{Blackshear93}{Blackshear, P. J. 1993.}  {The MARCKS family
    of cellular protein kinase C substrates.  \textit{J. Biol. Chem.}
    268:1501-1504. }
  
\bibitem{Czech2000}{Czech, M. P. 2000.}  {PIP2 and PIP3: complex
    roles at the cell surface.  \textit{Cell.} 100:603-606. }
  
\bibitem{Dan93}{Dan, N., P. Pincus, and S. A. Safran. 1993.}
  {Membrane-induced interactions between inclusions.
    \textit{Langmuir.} 9:2768-2771. }

  
\bibitem{Evans94}{Evans, D. F., and H. Wennerstrom. 1994.} {The
    colloidal domain, where physics, chemistry, biology, and technology
    meet, 2nd Ed. VCH publishers, New York.}
  
\bibitem{Fleck02}{Fleck, C., R. R. Netz, and H. H. von Gr\"unberg.
    2002.}  {Poisson-Boltzmann theory for membranes with mobile
    charged lipids and the pH-dependent interaction of a DNA molecule
    with a membrane \textit{Biophys. J.} 82:76-92.}

\bibitem{Gilson95}{Gilson, M. K. 1995. }  {Theory of electrostatic
    interactions in macromolecules. \textit{Curr.  Opin. Struct.
      Biol.} 5:216-223.}



\bibitem{Groves97}
{\emph{Groves, J. T., S. G. Boxer, and H. M. McConnell. 1997.}}
{\emph{Electric field-induced reorganization of two-component
supported bilayer membranes. \textit{Proc. Natl. Acad. Sci.} 
94:13390-13395.}}

\bibitem{Groves98}
{Groves, J. T., S. G. Boxer, and H. M. McConnell. 1998.}
{Electric field-induced critical demixing in lipid bilayer membranes. 
\textit{Proc. Natl. Acad. Sci.} 
95:935-938.}


\bibitem{Harries98}
{\emph{Harries, D., S. May, W. M. Gelbart, and A. Ben-Shaul. 1998.}}
{\emph{Structure, stability, and thermodynamics of lamellar DNA-lipid 
complexes. \textit{Biophys. J.} 75:159-173.}}




\bibitem{Heimburg99}{Heimburg, T., B. Angerstein, and D. Marsh. 1999.}
  {Binding of peripheral proteins to mixed lipid membranes: effect of
    lipid demixing upon binding.  \textit{Biophys. J.} 76:2575-2586.}
  
\bibitem{Honig93}{Honig, B., K. Sharp, and A. Suei-Yang. 1993.}
  {Macroscopic models of aqueous solutions: biological and chemical
    applications.  \textit{J. Phys. Chem.} 97:1101-1109.}
  
\bibitem{Honig95}{Honig, B., and A. Nicholls. 1995.}  {Classical
    electrostatics in biology and chemistry.  \textit{Science.}
    268:1144-1149.}
  
\bibitem{Katan97}{Katan, M., and R. L. Williams. 1997.}
  {Phosphoinositide-specific phospholipase C: structural basis for
    catalysis and regulatory interactions.  \textit{Semin. Cell. Dev.
      Biol.} 8:287-296.}


\bibitem{Kleinschmidt97}{Kleinschmidt, J. H., and D. Marsh. 1997.} {
    Spin-label electron spin resonance studies on the interactions of
    lysine peptides with phospholipid membranes.  \textit{Biophys. J.}
    73:2546-55.}

\envThesisRemark{\bibitem{Lau99}{Lau, A. W. C., and P. Pincus. 1999.} 
    {Binding of oppositely charged membranes and membrane 
    reorganization \textit{Eur. Phys. J.} 10:175-180.}}
  
\bibitem{Laux00}{Laux, T., K. Fukami, M. Thelen, T. Golub, D. Frey,
    and P. Caroni. 2000.}{GAP43, MARCKS, and CAP23 modulate
    PI(4,5)P(2) at plasmalemmal rafts, and regulate cell cortex actin
    dynamics through a common mechanism. \textit{J. Cell. Biol.}
    149:1455-72.}


\bibitem{Lee94}
{\emph{Lee, K. Y. C., J. F. Klinger, and H. M. McConnell. 1994.}}
{\emph{Electric-field-induced concentration gradients in lipid monolayers.
\textit{Science.} 263:655-658.}}

\bibitem{Lee95}
{\emph{Lee, K. Y. C., and H. M. McConnell. 1995.}}
{\emph{Effect of electric-field gradients on lipid monolayer membranes.
\textit{Biophys. J.} 68:1740-1751.}}


\bibitem{Lemmon03}{Lemmon, M. A. 2003.} {Phosphoinositide recognition
    domains. \textit{Traffic.} 4:201-213.}
  
\bibitem{Liu98}{Liu, Y., L. Casey, and L. J. Pike. 1998.}
  {Compartmentalization of phosphatidlyinositol 4,5- bisphosphate in
    low-density membrane domains in the absence of caveolin.
    \textit{Biochem. Biophys.  Res. Comm.} 245:684-690.}
  
\bibitem{Matsuoka00}{Matsuoka, Y., X. Li, and V. Bennett. 2000.}
  {Adducin: structure, function and regulation. \textit{Cell. Mol.
      Life Sci.} 57:884-95.}
  
\bibitem{May00}{May, S. 2000.} {Theories on structural perturbations
    of lipid bilayers.  \textit{Curr. Opin. Coll. Interface Sci.}
    5:244-249.}
                                               
\bibitem{Avinoam00}{May, S., D. Harries, and A. Ben-Shaul. 2000.}
  {Lipid demixing and protein--protein interactions in the adsorption
    of charged proteins on mixed membranes. \textit{Biophys. J.}
    79:1747-1760.}
  
\bibitem{Avinoam02}{May, S., D. Harries, and A. Ben-Shaul. 2002.}
  {Macroion-induced compositional instability of binary fluid
    membranes.  \textit{Phys. Rev. Lett.} 89:268102-1-4.}
    
\bibitem{Mclauglin95}{McLaughlin, S., and A. Aderem. 1995.}  {The
    myristoyl-electrostatic switch: a modulator of reversible
    protein--membrane interactions.  \textit{Trends Biochem. Sci.}
    20:272-276.}
  
\bibitem{Mclauglin02}{McLaughlin, S., J. Wang, A. Gambhir, and D.
    Murray. 2002.}  {PIP$_2$ and proteins: interactions,
    organization, and information flow.  \textit{Annu. Rev. Biophys.
      Biomol. Struct.} 31:151-175.}
  
\bibitem{MOE}{MOE} {Copyright \copyright 1997-2002 Chemical computing
    group info@chemcomp.com}
  
\bibitem{Murray97}{Murray, D., N. Ben-Tal, B. Honig, and S.
    McLaughlin. 1997.}  {Electrostatic interaction of myristoylated
    proteins with membranes: simple physics, complicated biology.
    \textit{Structure.} 5:985-989.}
  
\bibitem{Murray99}{Murray, D., A. Arbuzova, G. Hangyas-Mihalyne, A.
    Gambhir, N. Ben-Tal, B. Honig, and S. McLaughlin. 1999.}
  {Electrostatic properties of membranes containing acidic lipids and
    adsorbed basic peptides: theory and experiment.  \textit{Biophys.
      J.} 77:3176-3188.}

\bibitem{Murray01}{Murray, D., S. McLaughlin, and B. Honig. 2001.}
  {The role of electrostatic interactions in the regulation of the
    membrane association of G protein beta gamma
    heterodimers. \textit{J. Biol. Chem.} 276:45153-45159.}
  
\bibitem{Murray02}{Murray, D., A. Arbuzova, B. Honig, and S.
    McLaughlin. 2002.}  {The role of electrostatic and nonpolar
    interactions in the association of peripheral proteins with
    membranes. \textit{Curr. Top. Membr.} 52:271-302.}
  
\bibitem{MurrayAndHonig02}{Murray, D., and B. Honig. 2002.}
  {Electrostatic control of the membrane targeting of C2 domains.
    \textit{Mol.  Cell.} 9:145-54.}
    

\bibitem{Netz01}{\emph{Netz, R. R. 2001.}}
{\emph{Electrostatistics of counter-ions at and between planar charged
walls: From Poisson-Boltzmann to the strong-coupling theory.
\textit{Eur. Phys. J. E} 5:557-574.}}


  \envThesisRemark{ 
  \bibitem{Ohshima74A}{Ohshima, H. 1974.}  {Diffuse
      double-layer interaction between 2 parallel plates with constant
      surface charge-density in an electrolyte solution. I. The
      interaction between similar plates. \textit{Colloid and
        Polymer Science} 252:158-164.} 
  \bibitem{Ohshima74B}{Ohshima, H. 1974.}  {Diffuse
      double-layer interaction between 2 parallel plates with constant
      surface charge-density in an electrolyte solution. II. The
      interaction between dissimilar plates. \textit{Colloid and
        Polymer Science} 252:257-267.} 
  \bibitem{Ohshima75}{Ohshima, H. 1975.}  {Diffuse
      double-layer interaction between 2 parallel plates with constant
      surface charge-density in an electrolyte solution. III. Potential
      energy of double layer interaction. \textit{Colloid and
        Polymer Science} 253:150-157.} }

\bibitem{Parsegian72}{Parsegian, V. A., and G. David. 1972.}  {On the
    electrostatic interaction across a salt solution between two
    bodies bearing unequal charges. \textit{Biophys. J.}
    12:1192-1204.}
  
\bibitem{Payrastre01}{Payrastre, B., K. Missy, S. Giuriato, S. Bodin,
    M. Plantavid, and M. Gratacap. 2001.}  {Phosphoinositides: key
    players in cell signalling, in time and space. \textit{Cell.
      Signal.} 13:377-387.}
    
  
\bibitem{Cafiso02}{Rauch, M. E., C. G. Ferguson, G. D. Prestwich, and
    D. Cafiso. 2002.} {Myristoylated alanine-rich C kinase substrate
    (MARCKS) sequesters spin-labeled phosphatidylinositol
    4,5-bisphosphate in lipid bilayers. \textit{J. Biol. Chem.}
    277:14068-14076.}
  
\bibitem{Resh99}{Resh, M. D. 1999.} {Fatty acylation of proteins: new
    insights into membrane targeting of myristoylated and
    palmitoylated proteins. \textit{Biochim. Biophys. Acta.}
    1451:1-16.}
  
\bibitem{Rossi99}{Rossi, E. A., Z. Li, H. Feng, and C. S. Rubin.
    1999.}  {Characterization of the targeting, binding, and
    phosphorylation site domains of an A kinase anchor protein and a
    myristoylated alanine-rich C kinase substrate-like analog that are
    encoded by a single gene. \textit{J. Biol. Chem.} 274:27201-10.}


\bibitem{Rouzina96}
{\emph{Rouzina, I., and V. A. Bloomfield. 1996.}}  
{\emph{Competitive
electrostatic binding of charged ligands to polyelectrolytes: planar
and cylindrical geometries.
\textit{J. Phys. Chem.} 100:4292-4304.}}

\bibitem{Russ03}
{\emph{Russ, C., T. Heimburg, and H. H. von Gr\"unberg. 2003.}}
{\emph{The effect of lipid demixing on the electrostatic interaction of 
planar membranes across a salt solution. 
\textit{Biophys. J.} 84:3730-3742.}}

  
\bibitem{SimonsenA01}{Simonsen, A., A. E. Wurmser, S. D. Emr, and H.
    Stenmark. 2001.} {The role of phosphoinositides in membrane
    transport. \textit{Curr. Opin. Cell. Biol.} 13:485-492.}
  
\bibitem{Stauffer98}{Stauffer, T. P., S. Ahn, and T. Meyer. 1998.}
  {Receptor-induced transient reduction in plasma membrane
    PtdIns(4,5)P2 concentration monitored in living cells.
    \textit{Curr. Biol.} 8:343-346.}
  
\bibitem{Wang01}{Wang, J., A. Arbuzova, G. Hangyas-Mihalyne, and S.
    McLaughlin. 2001.} {The effector domain of myristoylated
    alanine-rich C kinase substrate binds strongly to
    phosphatidylinositol 4,5-bisphosphate.  \textit{J. Biol. Chem.}
    276:5012-5019.}
  
\bibitem{Wang02}{Wang, J., A. Gambhir, G. Hangyas-Mihalyne, D. Murray,
    U. Golebiewska, and S. McLaughlin. 2002.}  {Lateral sequestration
    of phosphatidylinositol 4,5-bisphosphate by the basic effector
    domain of myristoylated alanine-rich C kinase substrate is due to
    nonspecific electrostatic interactions. \textit{J. Biol. Chem.}
    277:34401-34412.}
  

\bibitem{Zhang2003}{Zhang, W., E. Crocker, S. McLaughlin, and S. O.
    Smith. 2003.}{Binding of peptides with basic and aromatic residues
    to bilayer membranes: phenylalanine in the myristoylated
    alanine-rich C kinase substrate effector domain penetrates into
    the hydrophobic core of the bilayer. \textit{J. Biol. Chem.}
    278:21459-21466.}
  
  

\end{thebibliography}
\end{document}